\documentstyle[11pt,aaspp4,psfig]{article}

\def\rxte{{\it RXTE}}

\begin{document}

\lefthead{Miller}
\righthead{290 Hz Spin Frequency in 4U~1636--536}

\title{Evidence for Antipodal Hot Spots During X-ray Bursts
From 4U 1636-536}

\author{M.\ Coleman Miller}
\affil{Department of Astronomy and Astrophysics, University of Chicago\\
       5640 South Ellis Avenue, Chicago, IL 60637, USA\\
       miller@bayes.uchicago.edu}
\authoremail{miller@bayes.uchicago.edu}

\begin{abstract}
The discovery of high-frequency brightness oscillations in
thermonuclear X-ray bursts from several neutron-star low-mass 
X-ray binaries has important implications for the beat frequency
model of kilohertz quasi-periodic brightness oscillations,
the propagation of 
nuclear burning, the structure of the subsurface magnetic fields
in neutron stars, and the equation of state of high-density
matter.  These implications depend crucially on whether the
observed frequency is the stellar spin frequency or its first
overtone.  Here we report an analysis of five bursts from
4U~1636--536 which exhibit strong oscillations at $\sim$580~Hz.
We show that combining the data from the first 0.75 seconds of
each of the five bursts yields
a signal at 290~Hz that is
significant at the $4\times 10^{-5}$ level when the number of
trials is taken into account.  
This strongly indicates that 290~Hz is the
spin frequency of this neutron star and that $\sim$580~Hz is its first
overtone, in agreement with other arguments about
this source but in contrast to suggestions in the literature
that 580~Hz is the true spin frequency.  The method used here,
which is an algorithm for combining time series data from the five bursts
so that the phases of the 580~Hz oscillations are aligned,
may be used in any source to search for weak oscillations that have
frequencies related in a definite way to the frequency of a strong
oscillation.

\end{abstract}


\section{INTRODUCTION}

Prior to the launch of the {\it Rossi} X-ray Timing Explorer
(\rxte) in December 1995, many theorists argued that
thermonuclear (Type 1) X-ray bursts on neutron stars are
almost certainly caused by ignition at a single point followed by
the spread of nuclear burning around the star, as opposed to simultaneous
ignition over the entire stellar surface (e.g., Fryxell
\& Woosley 1982; Nozakura et al. 1984; Bildsten 1995 ).
The
discovery with \rxte\ of high-frequency ($\sim 300-600$~Hz), 
nearly coherent high-amplitude brightness oscillations during Type~1 bursts
from several sources is of great significance because it
provides for the first time a wealth of observational data on the
propagation of nuclear burning in dense matter, which 
has implications for all phenomena involving such 
propagation, including novae and Type~Ia supernovae as well as
X-ray bursts.
The properties of these oscillations
may also yield important information about the strength and nature of
subsurface magnetic fields in these stars, and even about the equation
of state of the dense matter in their cores.
The conclusions drawn depend strongly on whether the frequency of the
burst oscillation from a particular source is the stellar
spin frequency or its first overtone.

The brightness oscillations during bursts from a given source
always have approximately the same frequency.  There are
six sources for which such oscillations have been reported (see
Strohmayer, Swank, \& Zhang 1998a for a more detailed summary):
4U~1728--34 (363~Hz), 4U~1636--536 (580~Hz), 4U~1702--43 (330~Hz),
KS~1731--260 (526~Hz), Aql~X-1 (549~Hz), and X~1743--29 (589~Hz).
In three of these sources oscillations have been observed in 
several bursts: 4U~1728--34 has had 17 such bursts,
4U~1636--536 has had five, and X~1743--29 has had three.  In no
burst from any source has an oscillation been detected at any other
frequency.  The extreme stability and coherence
of these burst oscillations provide compelling evidence
that the burst oscillation frequency is the stellar spin
frequency of the neutron star or an overtone
(see Strohmayer, Zhang, \& Swank 1997; Strohmayer
et al.\ 1998b).  It is thought that the oscillations themselves are
produced by the rotation of one or two ``hot spots" on the star that
are brighter than the surrounding surface.

The source 4U~1636--536 is a low-mass X-ray binary (LMXB) with an
orbital period of 3.8 hr (see, e.g., Pederson, van Paradijs, \&
Lewin 1981).  The beat-frequency model of the kilohertz 
quasi-periodic brightness oscillations (QPOs)
observed in the accretion-powered  X-ray emission from this
source predicts that the spin frequency of the neutron star
is roughly 290~Hz (see Miller, Lamb, \&
Psaltis 1998).  However, the power spectra of each of the five bursts
from 4U~1636--536 show a significant peak only at $\sim$580~Hz,
and as a result it has been suggested
that the spin frequency is 580~Hz (Strohmayer et al.\ 1998b). 
The $\sim$0.1c surface velocity implied by a 580~Hz spin frequency
is expected to generate significant power at the 1160~Hz first
overtone because of Doppler shifts and aberration (see Miller \&
Lamb 1998), but this has not yet been observed.
 
If instead
580~Hz is the first overtone of the spin frequency, as required
by the beat-frequency model, then 
in this source we must be seeing enhanced X-ray emission from two 
similar and nearly
antipodal hot spots, which are therefore likely to be the
locations of fuel accumulation, e.g., due to magnetic funneling.
In addition, the accumulated fuel must be confined to a relatively small
surface area.  The dipolar magnetic fields of $\sim 10^8-10^9$~G
inferred for these sources from spectral modeling 
(e.g., Psaltis, Lamb, \& Miller
1995) and from the sonic-point model of kilohertz QPOs (Miller 
et al.\ 1998) are insufficient  to confine the accreted fuel at the
depths where
nuclear burning occurs, which implies that the subsurface fields
are much stronger, perhaps $\sim 10^{11}$~G or higher.

If the strong 580~Hz oscillation observed in 4U~1636--536 can be
shown to be the first
overtone of the spin frequency, this would also have important implications 
for the compactness of neutron stars and their high-density equation
of state.  The more compact a neutron star is (i.e., the larger
its ratio of mass to radius), the greater the deflection of light
near its surface and hence the larger an area on the star that is
visible at infinity.  This implies in turn that the maximum
modulation in brightness produced by a rotating hot spot is lower
when the neutron star is more compact, and therefore that  an
observed amplitude of modulation places an upper limit on the
compactness (see Strohmayer 1997; Miller \& Lamb 1998).   These
limits are much stronger for a given oscillation amplitude  if
there are two hot spots instead of one (Miller \& Lamb 1998). 
Another difference of a 290~Hz spin frequency compared to a
580~Hz spin frequency is that it implies much lower surface
velocities, and hence a much smaller amount of power at 1160~Hz
due to Doppler shifts and aberration (Miller \& Lamb 1998).
It is therefore essential to analyze the data as sensitively as
possible to determine if there is any detectable oscillation
at 290~Hz in addition to the strong oscillation at 580~Hz.

Here we present a new analysis that combines data from five bursts
from 4U~1636--536.  Our approach is to use the strong signal at
580~Hz visible in all five bursts to combine the five time
series datasets
so that the signal at 580~Hz is maximized.  This procedure is
somewhat similar to the technique used by M\'endez et al.\ (1998)
to detect weak signals in the persistent emission from
4U~1608--52, except that they shifted and added power spectra
to maximize a strong peak (and hence added the data incoherently)
whereas we combine the time series data coherently, which improves the
sensitivity significantly.  We then analyze the initial
second of the combined burst data, which is the interval during
which any signal at 290~Hz 
is expected to be strongest, and find a prominent peak at
290~Hz.  The significance
of the peak at 290~Hz is $4\times 10^{-5}$ when the number of
trials is taken into account.  Therefore,
a significant signal is present at 290~Hz in one or more of the bursts, 
implying that 
290~Hz is the true spin frequency.  Data from other bursts from
4U~1636--536, when combined in the same way, are expected to enhance the
signal.  This technique of combining time series data 
in phase, using a strong signal,
can also be used to search for weak oscillations in other sources,
such as
4U~1728--34 with its 17 bursts exhibiting brightness oscillations.

In \S~2 we describe the data analysis method that we use
and our results in more detail.
In \S~3 we discuss the implications of the results and summarize our
conclusions.

\section{METHOD AND RESULTS}

We acquired public-domain data from the High Energy Astrophysics
Science  Archive Research Center archives for five bursts from
4U~1636-536, which occurred on 28 December 1996 at 22:39:24 UTC,
28 December 1996 at 23:54:03 UTC, 29 December 1996 at 23:26:47
UTC, 31 December 1996 at 16:39:28 UTC, and  23 February 1997 at
09:42:49 UTC.   We used the data in Event Mode and analyzed all
of the counts together (i.e., we did not select by energy
channel).  We focused on data in the initial portion of the
bursts, because if the oscillations are caused by the rotation
of expanding hot spots (see, e.g., Strohmayer et al.\ 1997) it
is expected that harmonics and subharmonics of the frequency
of the primary peak
will be strongest near the start of the burst.
For each burst,
we therefore constructed a time series of countrate data $C_i(t)$,
for the $i$th burst, starting at the beginning of the burst as
determined by when the countrate rises above the persistent
countrate.

Observations of 4U~1636-536 and other X-ray bursters show that
the frequency of the strong peak, in this case near 580~Hz, 
can vary 
in a complicated way that is different from burst to  burst, with
frequency changes of a few Hertz in less than a second
(Strohmayer et al.\ 1998a).   The power at the primary
$\sim$580~Hz peak is therefore spread significantly unless
the temporal variation of this frequency 
is modeled
carefully and separately for each burst (see, e.g., 
Strohmayer et al.\ 1998a).  Since the
existence of the $\sim$580~Hz peak is secure, models of its
frequency variation can be explored without
increasing the number of trials.

After extensive exploration of functional forms for the
frequency behavior, we find that a sufficiently descriptive
model has five parameters
for each burst: frequency and frequency derivative for an
initial, short time; a different frequency and frequency
derivative for the remainder of the data; and a break time
between the two.  This is similar to the frequency behavior reported for
the brightness oscillations during bursts from several other
sources (Strohmayer et al.\ 1998a).  Given this relatively simple description,
there is a frequency jump at the break time, which is not
expected to be present in a more accurate representation of
the frequency behavior.
The break time was searched over the range
from 0.125~s to 0.625~s, and the frequency was constrained so that
nowhere in the interval was it less than 576~Hz or greater than
585~Hz, to exclude spurious solutions produced by noise in the
power spectrum.  The model that we use for the behavior of the 
frequency during a particular burst is
therefore
\begin{eqnarray}
\omega(t)& =\omega_1+{\dot\omega}_1 t,\qquad  t<t_{\rm break}\\
& =\omega_2+{\dot\omega}_2 t,\qquad  t\geq t_{\rm break}\; .
\end{eqnarray}
These parameters are then varied for each burst to maximize
the power at 580~Hz.  In the next step, the data for
the five bursts are time-shifted with respect to each other
by times $\delta t_i<1/580$~s, so that
the power
\begin{equation}
P=\left| \int_0^T
\Sigma_i C_i(t+\delta t_i)e^{i\omega_i(t)}dt\right|^2
\end{equation}
of the combined data is maximized at 580~Hz.  We expect that
this is equivalent to aligning the phases of the strong
oscillation in the bursts.  Here $\omega_i(t)$
is the frequency fit for the $i$th burst defined by
equations (1) and (2) above, and
without loss of generality we can set $\delta t_1=0$.
The time $T$ is the time interval of the data included;
in general this time could be different for each burst, but
for simplicity we integrate over the same interval for all
five bursts.  In the final
step, $T$ is varied so as to maximize the 580~Hz power.  This
maximum occurs for $T$=0.734~s.  The fits and the
resulting characterization of the burst brightness oscillations 
in 4U~1636--536 will be
described in detail in a forthcoming paper.

In total, we use 30 parameters to describe the frequency
behavior of the $\sim$580~Hz peak in all five bursts: five for each of
the five bursts, four for the relative phases between bursts, and 
the integration time.  We emphasize again that the strength
of this peak allows us to do such modeling without introducing
any trials.

 \begin{figure*}[t]
 \vbox{\vskip-0.2truein\hbox{\hskip 1.2truein
 \psfig{file=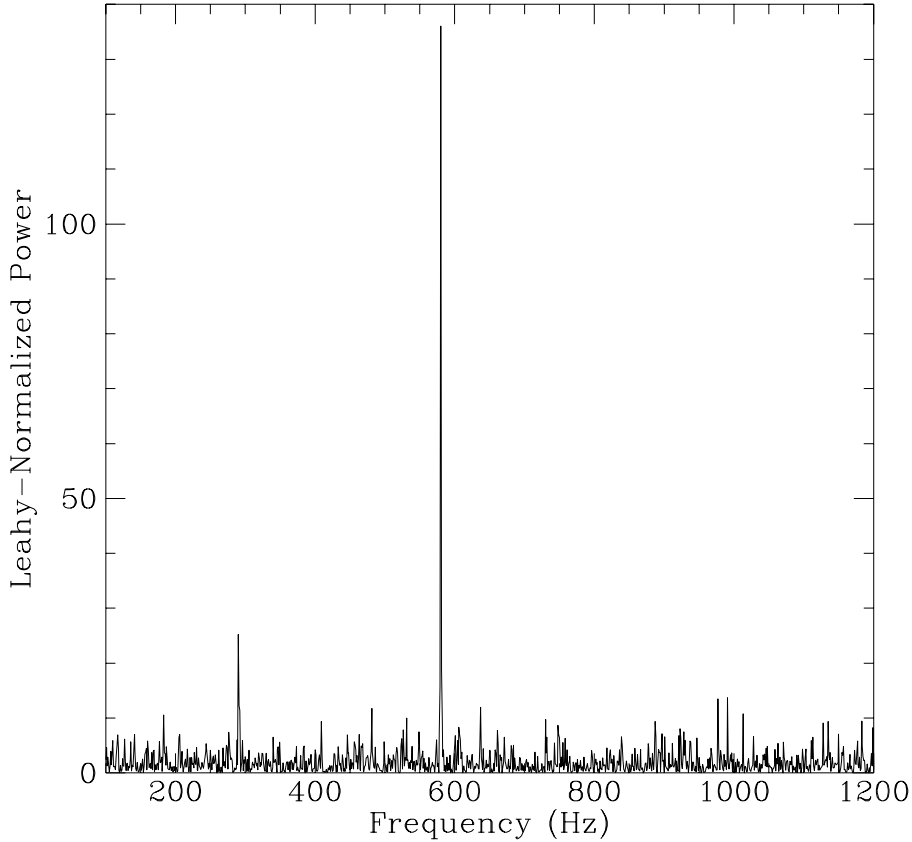,height=4.8truein,width=4.8truein}}}
 \vskip-0.2truein
 \caption[fig1]{
Leahy-normalized power spectra above 100~Hz for the first
0.734~s of the combined data set of the five bursts.  The
probability of a peak in this graph occurring by chance is
just $\exp(-P/2)$, where $P$ is the power of the peak.  The
power at 580~Hz is 136 and that at 290~Hz is 25.6; the latter
has a chance probability of $2.8\times 10^{-6}$.  Taking into
account the 16 trials required to phase-align the
290~Hz peaks in the five bursts, the significance of the peak
at 290~Hz is $4.4\times 10^{-5}$.}
 \end{figure*}

This procedure maximizes the power of the signal at 580~Hz and will
also maximize the power of the signal at any integral multiple
of 580~Hz (e.g., at the 1160~Hz overtone of this frequency).
However, it is not guaranteed to maximize the phase coherence
at subharmonics such as 290~Hz.  For example, if the 580~Hz oscillations
for two different bursts are in phase, then the 290~Hz oscillations
are either in phase or $\pi$ radians out of phase.  Therefore,
after maximizing the signal at 580~Hz in two bursts as above, it is necessary
to determine which of the two possible relative oscillation
phases maximizes the
signal at 290~Hz.  For five bursts there are four such choices
of relative phase, and hence there are $2^4$=16 trials.  The
difference between the power produced by the two different phase 
choices in each of the four steps is so great that there are no
ambiguities. 

The 100--1200~Hz power spectrum constructed from the time series
data combined in this way is shown in Figure~1.  An
indication of the success of this method in bringing out the power
at the 580~Hz peak is that if a power spectrum is constructed for
the initial 0.734 seconds of the individual bursts, and the frequency
model is simply a constant frequency, then the highest Leahy et al.\ (1983)
normalized
power near 580~Hz for any burst is 37 whereas the power at
580~Hz in Figure~1 is 136.  This difference occurs because of the
large variation in the frequency of the oscillation near 580~Hz, which becomes
rapidly dephased compared to any constant frequency signal.
This figure also shows a strong peak
at 290~Hz, with a Leahy-normalized power of 25.6.  This means that,
in the combined data, the amplitude at $\sim$290~Hz is only a factor
of 2.3 less than the amplitude at $\sim$580~Hz, and hence there
can actually be a substantial amount of asymmetry between the
two main emitting spots.  Taking into account
the 16 trials performed for the 290~Hz peak, the significance of
the peak is $4.4\times 10^{-5}$.  This significance estimate is
likely to be conservative, because in order to bypass issues
of multiple trials we have assumed not only that the $\sim$290~Hz
signal is always at exactly half the frequency of the $\sim$580~Hz
signal, but also that the relative phase between the 580~Hz 
oscillation and the
290~Hz oscillation is constant throughout a burst and is the same from
burst to burst.  In reality, the relative phase is not likely to
be constant, and hence the true power at 290~Hz is probably 
greater than estimated here.

\section{DISCUSSION AND CONCLUSIONS}

The very significant peak at 290~Hz indicates that this is the spin
frequency of the neutron star in 4U~1636--536, and that 580~Hz
is the first overtone.  Several major consequences follow:

\noindent (1)~From the strength of the 580~Hz signal compared to
the 290~Hz signal, there are two very similar and nearly antipodal
hot spots on the surface, and they are almost equally visible
to us.  This almost certainly implies that accreting gas is
being funneled to the two hot spots by an external magnetic field.
In
addition, either the two spots or our line of sight must be 
nearly in the rotational equator, because otherwise
we would see much stronger emission from one spot than from the 
other, and hence a strong signal at 290~Hz.  Of the two possibilities
it is more likely that it is the hot spots that are near the rotational
equator, because if our line of sight were close to the rotational
equator then it would probably be blocked by the accretion disk.
Hence, this source is probably close to being an orthogonal
rotator, which has interesting implications for the evolution of
its magnetic field geometry. 

\noindent (2)~The rapidity with which the signal at the 580~Hz
first overtone of the spin frequency appears
indicates that thermonuclear ignition must be communicated
quickly from one pole to the other (F.~K. Lamb, personal
communication).  Our analysis of the first
burst in our sample shows that there is a significant oscillation
at 580~Hz within 0.03 seconds of the onset of this burst.
The distance between poles
is approximately $3\times 10^6$~cm, so the required velocity
is in excess of $10^8$~cm~s$^{-1}$.  This velocity is
significantly greater than even the largest velocities
$\sim 3\times 10^7$~cm~s$^{-1}$
estimated for deflagration waves.  Hence, if the ignition at one
pole is communicated to the other pole via propagation of a
nuclear burning front, the front must be a detonation wave.

\noindent (3)~The existence of a strong signal at 580~Hz in the tail
of the burst (see Strohmayer et al.\ 1998c) indicates that
fuel is not only {\it funneled} toward the magnetic poles,
but is {\it confined} there before the start of the
X-ray burst.  If the fuel were not confined (e.g., if the
magnetic field were too weak to prevent the fuel from
spreading) then the fuel would spread almost evenly over the
entire surface, so that even if ignition occurs almost simultaneously
at the two poles, there would be no strong oscillation
at 580~Hz in the tail of the burst (F.~K. Lamb, personal communication).

\noindent (4)~The large amplitudes reported by Strohmayer et al.\ 
(1998c) near the beginning of one burst place strong constraints on
the compactness of the neutron star.  The maximum modulation
amplitude produced by rotation is much less when there are two
emitting spots than when there is only one (see Strohmayer 1997;
Miller \& Lamb 1998).  Indeed, the 75\%$\pm$17\%
peak-to-peak amplitude (50\% rms) at 580~Hz reported for a 1/16 
second
interval near the beginning of one burst from 4U~1636--536 
(Strohmayer et al.\ 1998c) is so large that Strohmayer et al.\ (1998c)
used it as an argument that only one pole was emitting and hence
that 580~Hz is the true spin frequency.  An amplitude this large
can, however,
be produced by two antipodal spots if bandwidth corrections and
surface beaming are taken into account (Miller \& Lamb 1998).
Moreover, the quoted amplitude applies to a countrate spectrum
from which an estimated background (calculated using the 20 seconds
prior to the burst) has been subtracted.  This implicitly assumes
that the background is
constant during the initial phase of the burst.  If this is not the case,
the true amplitude could be significantly
less.  Nonetheless, the existence of such a high amplitude from
two hot spots has excellent promise for constraining strongly
the compactness of this neutron star.

\noindent (5)~The establishment of 290~Hz as the spin frequency
of 4U~1636--536 provides further support for the beat-frequency model of
the kilohertz brightness oscillations detected in the
persistent accretion-powered X-ray emission from many neutron-star
LMXBs (Strohmayer et al.\ 1996; Miller et al.\ 1998;
see van der Klis 1998 for an observational overview of these
oscillations).  In the beat-frequency model, the spin frequency of
4U~1636--536 is predicted to be approximately 290~Hz (see Miller
et al.\ 1998), and therefore the detection of a signal at 290~Hz
strengthens confidence in the inferences drawn from the
beat-frequency model. For example, observations of
4U~1820--30 interpreted using this model provide good evidence 
for the detection of an innermost
stable circular orbit (a key prediction of strong-gravity general
relativity) and for the existence of a 2.2--2.3$M_\odot$ neutron
star (Zhang et al.\ 1998b), which have profound implications 
for our understanding of gravity and nuclear forces.

In conclusion, the results presented here show that the use of a
strong oscillation as a clock to align time series data
in some bursts is an
effective method to search for weaker oscillations at related frequencies.  
This procedure is
especially powerful when a single source has more than one burst
with a strong oscillation (e.g., 4U~1728--34, for which 17
bursts exhibiting brightness oscillations have 
been observed; see Strohmayer et
al.\ 1998a), because coherent addition of data is then possible.
It can also be used to improve signal detection during a single burst
(as, for example, was done by Zhang et al.\ 1998a using data from a
burst from Aql X-1).  The method may also be used to characterize
the properties of a peak whose existence has been demonstrated,
such as either the 580~Hz peak or the 290~Hz peak in
4U~1636--536.  In a future paper we will report in detail
the frequency and amplitude behavior of the
oscillations in this source. Such characterization and detection holds
outstanding promise as a sensitive probe of the properties of
nuclear burning in
X-ray bursts and of neutron stars themselves.

\acknowledgements

I am grateful to Don Lamb, Carlo Graziani, and Jean Quashnock for
their  helpful suggestions about the data analysis and its
evaluation.  I also appreciate discussions with Don Lamb and Fred
Lamb about the physics of bursts and the implications of the
oscillations, and for their comments on a previous version of this
paper.  Will Zhang provided the dates of the bursts used
in this analysis and the data for one of the bursts.  This
research has made use of data obtained through the High Energy
Astrophysics Science Archive Research Center Online Service,
provided by the NASA/Goddard Space Flight Center.  This work was
supported in part by NASA grant NAG~5-2868 and NASA AXAF contract
SV~464006.

\end{document}